# Discriminating Between Coherent and Incoherent Light with Metasurfaces

T. Frank,[1,2] O. Buchnev,[2,3] T. Cookson,[1] M. Kaczmarek,[1] P. Lagoudakis[1] and V. A. Fedotov[2,3]*
[1] Physics and Astronomy, University of Southampton, SO17 1BJ, UK
[2] Optoelectronics Research Centre, University of Southampton, SO17 1BJ, UK
[3] EPSRC Centre for Photonic Metamaterials, University of Southampton, SO17 1BJ, UK
*email: vaf@orc.soton.ac.uk

**Abstract:** Metasurfaces represent a powerful paradigm of optical engineering that enables one to control the flow of light across material interfaces. We report on a discovery that metallic metasurfaces of a certain type respond differently to spatially coherent and incoherent light, enabling robust speckle-free discrimination between different degrees of coherence. The effect has no direct analogue in conventional optics and may find applications in compact metadevices enhancing imaging, vision, detection, communication and metrology.

Over the last decade the concept of artificially engineered media (known as metamaterials) has revolutionized the field of optics, pushed the boundaries of microfabrication and stimulated the development of novel characterization techniques [1, 2]. Recent demonstrations of anomalous reflection and refraction of light by metasurfaces opened another exciting chapter in optical engineering [3]. Metasurfaces correspond to a class of low-dimensional (planar) metamaterials and are typically formed by optically thin metal films periodically patterned on a sub-wavelength scale. Despite their vanishing thickness, metasurfaces interact strongly with light, which they can transmit, absorb or reflect without diffraction, effectively acting as optical media of zero dimension in the direction of light propagation. That sets metasurfaces aside from diffractive resonant waveguide gratings (aka photonic crystal slabs) [4, 5] and perforated metal films exhibiting extraordinary optical transmission [6]. Metasurfaces are fully compatible with existing fabrication processes adopted by CMOS technology, and offer unmatched flexibility in the design and control of light propagation, replacing conventional bulk optical components and exhibiting exotic electromagnetic phenomena. In particular, metasurfaces have already enabled spectral [7, 8, 9] and directional [10, 11, 12] filtering, absorption enhancement and energy harvesting [13, 14, 15, 16, 17], polarization control [18, 19, 20] and analysis [21, 22], imaging [23, 24, 25] and sensing [26, 27], as well as have allowed the demonstration of exotic effects of asymmetric transmission [28, 29] and specular optical activity [30].

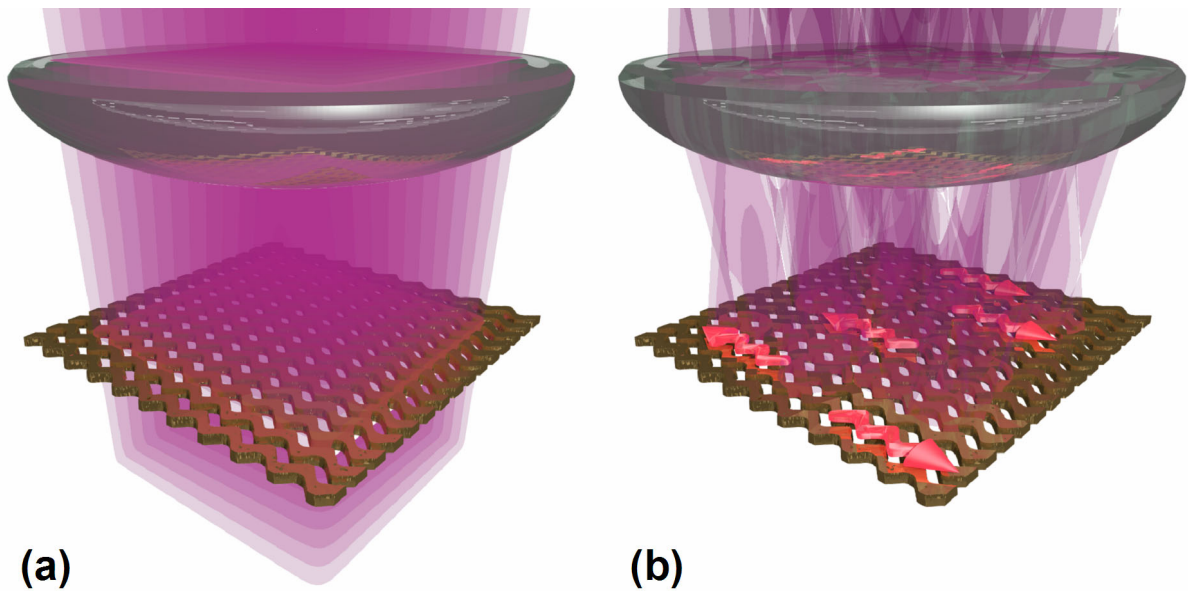

**Figure 1. Discriminating between spatially coherent and incoherent light with metasurfaces.** Panels (a) and (b) give artistic impressions of a non-diffractive metallic metasurface interacting with correspondingly coherent (a) and incoherent (b) light. Transmission under spatially incoherent illumination appears to be strongly suppressed via non-local light scattering, which is mediated by the framework of the metasurface.

In this paper we describe and investigate an intriguing optical phenomenon whereby apparently trivial, non-diffracting metallic metasurfaces exhibit different levels of speckle-free transmission, depending on whether they are illuminated with spatially coherent or incoherent light (as illustrated in Fig. 1). This effect, previously unseen in artificially engineered media, is robust and exceptionally strong, and does not affect the beam quality, which makes it immediately suitable for practical applications, such as optical metrology, imaging and communications.

The phenomenon was discovered experimentally with zigzag metasurfaces operating in the near-IR part of the spectrum. The metasurfaces were milled with a focussed ion beam in an 80 nm thick film of amorphous gold that had been sputtered on a 0.5 mm thick fused-quartz substrate beforehand. The fabricated samples featured complementary versions of the zigzag pattern (shown in Figs. 2a and 2b), which correspond to the arrays of continuous nanowires (ZZnW) and their inversion, i.e., continuous nanoslits (ZZnS). Both the nanowires and nanoslits had the width of approximately 80 nm. The unit cell of the zigzag pattern contained two straight segments of the nanowire (or nanoslit) forming a right angle and had the dimensions of 660 nm x 520 nm, which rendered the metasurfaces as non-diffracting in the near-IR. The fabricated samples had the area of 21.1 µm x 20.8 µm and encompassed a total of 1280 unit cells (see Fig. 2c).

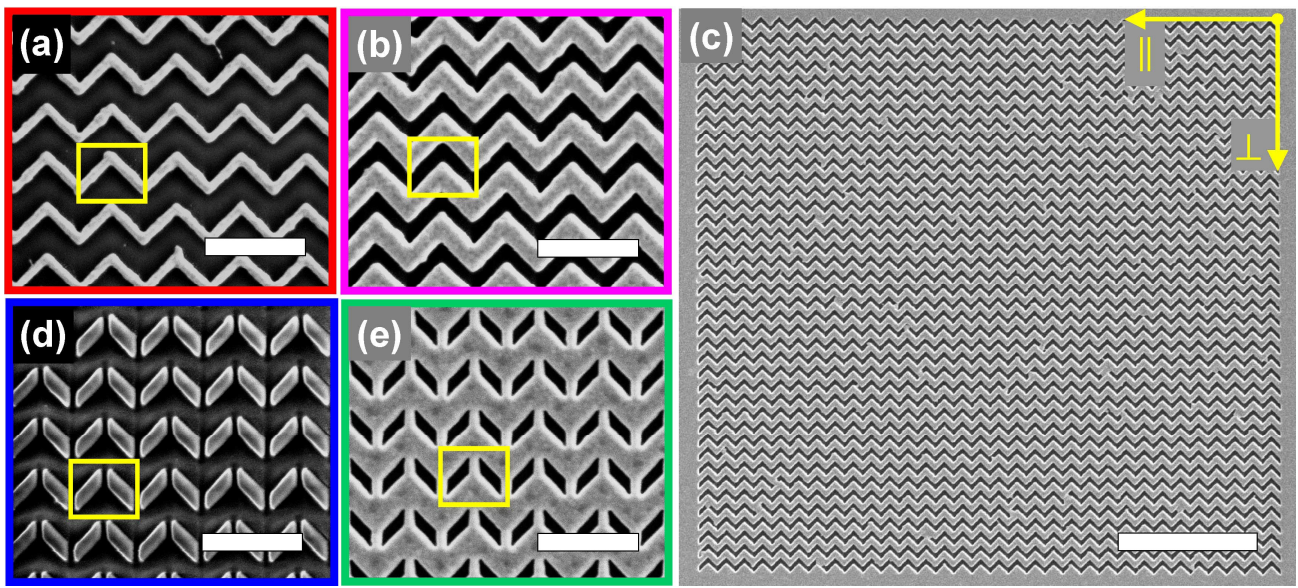

**Figure 2. Scanning electron micrographs of zigzag metasurfaces.** Images in panels (a) and (b) show fragments of the studied samples featuring continuous zigzag nanowires and nanoslits, respectively. Panel (c) provides a full view of the sample from panel (b). Images in panels (d) and (e) show fragments of the reference samples featuring broken zigzag nanowires and nanoslits, respectively. Yellow boxes highlight elementary unit cells of the metasurfaces. Scale bars correspond to 1 µm in panels (a), (b), (d) and (e), and to 5 µm in panel (c).

Optical properties of those metasurfaces were characterized in transmission at normal incidence using a near-IR microspectrophotometer equipped with an incandescent white-light source (WLS) and a broadband linear polarizer. For a ZZnW metasurface, the incident polarization was set perpendicular (⊥) to the rows of the zigzag pattern, while for a ZZnS metasurface the polarization was set in the direction parallel (∥) to the zigzag rows, as marked in Fig. 2c. The measured spectra, normalized to the transmission of a bare substrate, are shown by solid curves in Figs. 3a and 3b. In the wavelength range 0.9 – 1.5 µm both spectra featured, what appeared to be, a pair of resonances separated by a gap of about 0.2 µm. Given the complementarity of ZZnW and ZZnS patterns, the features of the two transmission spectra were also complimentary (as dictated by Babinet's principle), i.e. peeks in one spectrum corresponded to dips in the other, and vice versa. Intriguingly, the overall profile of the spectral response in each case resembled closely that of Fano resonances normally exhibited by metasurfaces with substantially more complex patterning [31]. More intriguingly, the results of our measurements seemed to disagree – even at the qualitative level – with the predictions of rigorous numerical modelling that informed the design of our metasurfaces (shown by dotted curves in Figs. 3a and 3b). Indeed, for the same wavelength range the modelled response in each case displayed a single resonance centred at $\lambda = 1.1$ µm.[†] It resulted from the excitation of the fundamental $\lambda/2$-current mode – the most common *localized* (dipolar) resonant mode that had been particularly favoured by microwave and RF antennas and planar metamaterials of various designs [7, 8] (including those patterned continuously akin to the zigzag metasurfaces [32, 33]). In the zigzag metasurfaces the mode corresponded to a standing wave of charge (plasma) oscillations, which built up *locally* in every straight segment of the pattern, once the half of the wave's period fitted the length of the segment (see insets to Figs. 3a and 3b). That rendered each segment acting as an independent half-wavelength strip (slit) resonator, which determined the response of the metasurfaces. The noted strong (and rather unexpected) discrepancy between the theory and experiment called for careful examination and verification of the methodology used, the accounts of which are given below.

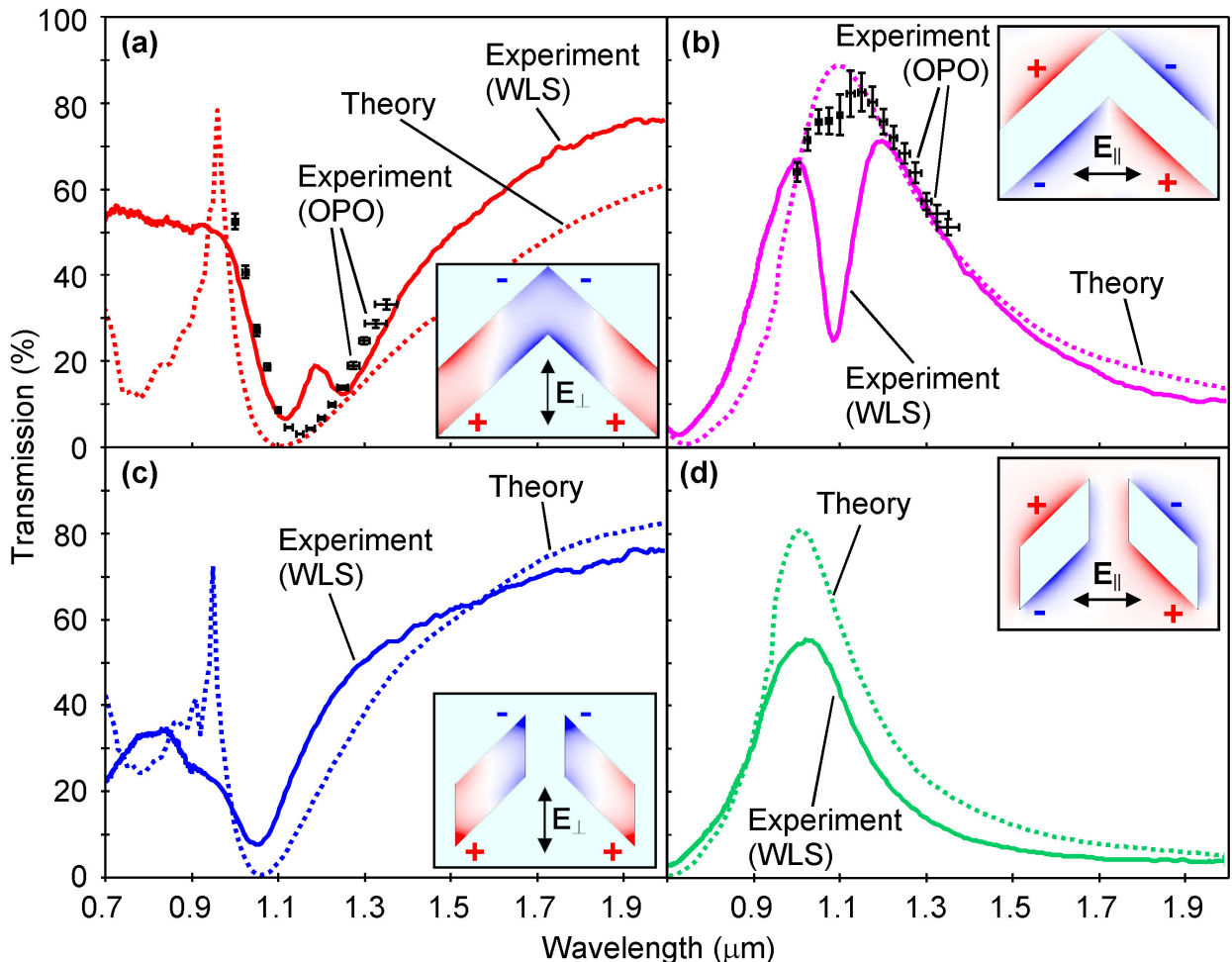

**Figure 3. Transmission response of zigzag metasurfaces.** Data in panels (a) and (b) characterize transmission spectra of the studied metasurfaces featuring continuous zigzag nanowires and nanoslits, respectively. Data in panels (c) and (d) characterize transmission of the reference metasurfaces formed by broken zigzag nanowires and nanoslits, respectively. Dashed curves show numerically modelled spectra, which informed the designs of zigzag metasurfaces. Solid curves display the data measured experimentally using linearly polarized light from an incandescent white light source (WLS). Crosses represent the data acquired experimentally using linearly polarized light from an optical parametric oscillator (OPO). The length of the bars in every cross indicates the uncertainty of the corresponding measurement. For each data set the polarization of incident light is shown in the respective inset along with the actual geometry of the unit cell. Light-blue sections of the unit cells correspond to air. The insets also reveal the current mode excited in each unit cell at the resonance by displaying a snapshot of the modelled distribution of charge density (colours).

---

[†] Features visible below 0.95 µm in the modeled spectrum of a ZZnW metasurface were the signs of diffraction occurring in the bulk of the substrate due to 660 nm period of the metasurface. It became possible to observe due to high overall transparency of the ZZnW metasurface at those wavelengths and the idealized configuration of the modeled sample.

For modelling transmission of the zigzag metasurfaces we used a well-established (for periodic structures) and computationally efficient approach, whereby the simulation domain ($< 4\lambda^3$) accommodated only one unit cell of the modelled metasurface. The opposite sides of the cell and, correspondingly, the faces of the domain contiguous to them were subjected to boundary conditions, periodic in $\perp$ and $\parallel$ directions. The remaining two faces of the domain (normal to the direction of light propagation) were terminated with perfectly matched layers (PMLs), with one of the faces set as a source of an electromagnetic wave. The extent of the domain along the propagation direction was greater than two wavelengths, which made sure that the unit cell of the metasurface was at least one wavelength away from either of the PMLs. With the above settings our model effectively described transmission of a normally incident plane wave through an infinitely large, periodically patterned metasurface sandwiched between (semi-infinite) slabs of air and fused quartz. The refractive index of the latter was set to 1.45 [34] and kept constant across the entire spectral range of interest. The dielectric function of gold was defined by the tabulated data [35]. The modelling was implemented with the help of the electromagnetics module of finite element method-based commercial simulation software COMSOL Multiphysics. Experimentally, optical characterization of the zigzag metasurfaces was carried out using a commercial microspectrophotometer developed by CRAIC Technologies on the basis of a ZEISS Axio microscope. It featured a cooled near-IR CCD array with spectral resolution of 0.8 nm and Köhler illumination system. The latter incorporated a tungsten-halogen lamp and was tuned to a x15 reflective objective with NA 0.28, producing broadband plane-wave illumination at near-normal incidence. The spectra were collected through a 22 μm x 22 μm aperture installed in the image plane of the microscope.

In all the previous works on nanostructured metasurfaces, where the above (or similar) methodology and instrumentation had been employed, a good agreement between the theory and experiment was reported (see, for example, [36, 37, 38, 39, 40, 41, 42, 43, 44, 45, 46, 47, 48, 49, 50, 51, 52]). The metasurfaces in those studies, however, had one feature in common – their designs were based on piecewise rather than continuous patterns. To judge on the importance of that observation we fabricated and subjected to the same testing procedure as above a set of *reference* samples, which resembled ZZnW and ZZnS metasurfaces with broken (i.e. piecewise) zigzags. Their pattern was derived from the original by introducing a 70 nm wide split in every corner of the zigzags, as shown in Figs. 2d and 2e. Importantly, since the resonant mode supported by the continuous pattern had its nodes located in the corners of the zigzags (see insets to Figs. 3a and 3b), the described modification did not affect the nature of the resonant response in the resulting nanostructures. This is evident from the insets to Figs. 3c and 3d, which show that the localized distributions of charge density sustained by the reference metasurfaces at their resonance were very similar to those calculated for the ZZnW and ZZnS metasurfaces (compare with the insets to Figs. 3a and 3b). Also, as in the case of continuous zigzags, the transmission spectra predicted for piecewise zigzags featured a single resonance spanning from 0.9 to 1.5 μm, though its centre appeared blue-shifted by about 50 nm due to physical shortening of the broken segments (compare dotted curves in Fig. 3). In spite of the apparent similarities between the two cases, we were able to reproduce experimentally the main features of the response predicted numerically for the reference samples (see Figs. 3c and 3d). Some quantitative mismatch did not affect the validity of the comparison, as it resulted from an uncertainty in specifying the dielectric function of gold and the difficulty of reproducing fine features of piecewise zigzags during the fabrication (e.g. sharp corners near the splits).[‡]

Our analysis, therefore, indicated that the discrepancy between the theory and experiment that had been observed earlier with the ZZnW and ZZnS metasurfaces corresponded to a genuine, previously unseen effect, somewhat exclusive to the continuous pattern. This discrepancy could only have resulted from the difference between the simulated and the actual illumination conditions, which thus far did not seem to matter and had been routinely disregarded in the study of metamaterials: the common modelling approach assumed the complete coherence of incident light (due to the harmonic field formulation and periodic boundary conditions), while the spectroscopic measurements involved spatially incoherent light (in our case, from a tungsten-halogen lamp). To verify our conclusion we re-measured the transmission spectra of the ZZnW and ZZnS metasurfaces using a combination of a wavelength-tuneable laser and broadband power meter. The laser source was a quasi-CW optical parametric oscillator (Chameleon Compact OPO) by Coherent with the tuning range 1.00 – 1.35 µm. Linearly polarized OPO output was focussed on the samples to a spot with the diameter of 50 µm by an achromatic 100 mm lens. Another 100 mm lens was used to collect the transmitted light and direct it towards the power meter. The spectra were acquired with the step of 25 nm and normalized to the transmission of a fused-quartz window (a rectangular opening that was milled in the gold film and had the same dimensions as the samples). The obtained data are plotted in Figs. 3a and 3b as crosses. Clearly, the predicted resonances were satisfactory reproduced by the experimental data, which confirmed that ZZnW and ZZnS metasurfaces could 'sense' the degree of light coherence, exhibiting in our case either 3-fold enhancement or suppression of their transmission under incandescent illumination.

---

[‡] As in Fig. 3a, the discrepancy below 0.95 μm in Fig. 3c corresponded to the diffraction regime in the bulk of the substrate due to 660 nm period of the metasurface and, therefore, was of no importance.

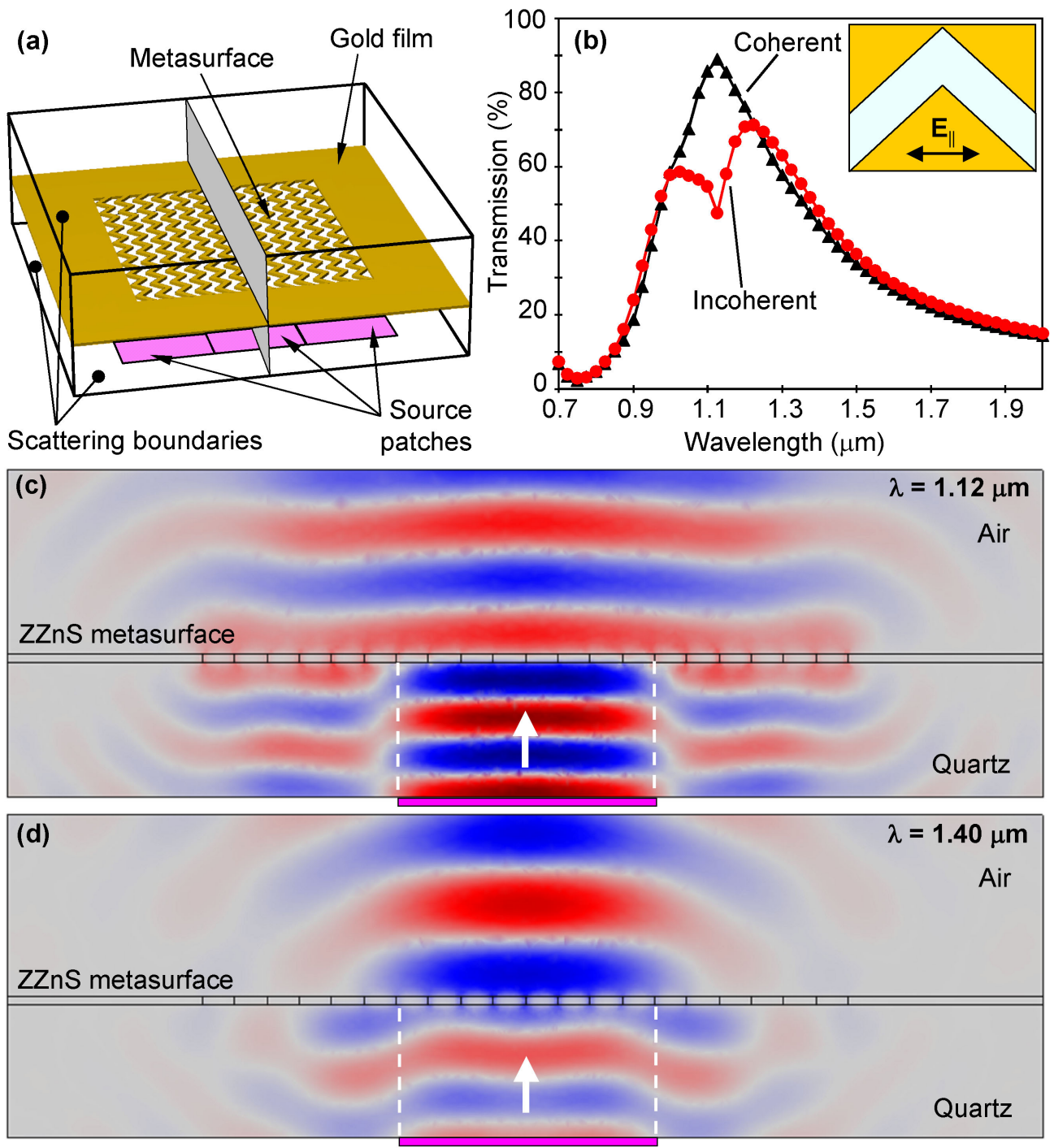

**Figure 4. Modelling response of zigzag metasurfaces to incoherent illumination.** Schematic in panel (a) depicts the layout of the computational domain. It encompasses a metasurface 6.6 x 6.2 μm² large, which is framed by 2 μm wide strip of unstructured gold film. All sides of the domain are set as scattering boundaries. The area directly underneath the metasurface is divided into nine squares with each square set to independently produce a light beamlet propagating upward. Plots in panel (b) represent calculated transmission spectra of the metasurface formed by continuous zigzag nanoslits (ZZnS). Data points corresponding to spatially incoherent illumination are shown with red solid circles. For comparison, the plots also include data points calculated for the case of coherent (i.e. plane wave) illumination using the same model, which are marked by black solid triangles. Colour maps in panels (c) and (d) capture the scattering of the wave of the central beamlet by ZZnS metasurface at the wavelengths of 1.12 μm and 1.40 μm, respectively. The maps are plotted for the cross-section splitting the computational domain in half along the zigzag rows, as shown in panel (a). The colours display the real part of the electric field of the wave, which is polarized in the plane of the cross-section. White arrows indicate the direction of incidence. White dashed lines indicate the lateral extent of the incident beamlet.

To understand the nature of the discovered phenomenon we have built a comprehensive computational COMSOL model that would enable us to simulate the response of the zigzag metasurfaces to incoherent light. The model featured a rather large simulation domain (> $46\lambda^3$), where we could faithfully reproduce the entire layout of the samples and, hence, avoid the use of periodic boundaries. Due to memory constrains of our computational hardware, the size of the modelled metasurfaces was limited to 10 x 12 unit cells (see Fig. 4a). Also, instead of PMLs, we used scattering boundaries all around – a robust alternative for large simulation domains, which helped us to ease the memory constraints and avoid the difficulty of formulating PMLs near the interfaces, where media appeared inhomogeneous [53]. One of the domain's faces parallel to the modelled metasurface contained nine square patches, each set to independently generate a light beamlet with the cross-section of 2.4 x 2.4 µm² (shown as purple in Fig. 4a). The dimensions of the cross-section matched the spatial coherence length of incandescent light ($d_c$) at $\lambda$ = 1.1 µm, which was estimated according to the expression $d_c = 3.832\,\lambda/2\pi\,N_A$ [54]. The other face parallel to the metasurface served as a detector capturing the overall power transmitted by the nanostructure in the forward direction. The spectral response to spatially incoherent illumination was given by the sum of the transmission spectra from nine separate runs of the model, each engaging only one particular beamlet. Figure 4b displays the resulting spectrum for the case of a ZZnS metasurface. It features a split resonance resembling very closely what was observed in the experiment under incandescent illumination. Some quantitative mismatch between the measured and calculated data is attributed to an uncertainty in specifying the dielectric function of amorphous gold, as well as to the limitations of our model. In particular, the coherence length in our simulations (as defined by the cross-section of the light beamlets) was kept constant across the entire wavelength range, while the modelled metasurface was smaller than the actual ZZnS sample by a factor of 10. To demonstrate that the profile of the calculated spectrum was not just an artefact of our modelling approach we compared it in Fig. 4b with the spectrum produced by the same model under fully coherent illumination (all nine beamlets were engaged at the same time).

The nature of the discovered effect and, more specifically, the origin of the split resonance can be deduced from the distribution of electric field near the ZZnS metasurface calculated at the split's centre wavelength. Such a distribution is plotted in Fig. 4c for the cross-section of the domain that divides the metasurface in half along the zigzag rows. It pictures a light wave polarized in the plane of the cross-section (∥-polarization), which emanates from the central patch at the bottom of the domain and propagates upward as a beamlet until scattered by the metasurface. While the incident wave is confined laterally to the area of the patch (exhibiting only minor spreading due to diffraction), the transmitted and reflected waves are seen to spread along the metasurface far beyond these confines. The resulting field configuration indicates that the mechanism of light scattering by the ZZnS metasurface is extremely non-local. Since metasurfaces do not diffract this mechanism must involve current (i.e. plasmon) waves, which due to continuity of the pattern can leak from a locally excited unit cell and propagate up and down the zigzags (as illustrated in Fig. 1b). In the ZZnS metasurface these waves are guided in the form of a mode confined to the nanoslits. It transports the excitation via the zigzag channels to other parts of the metasurface, where it is radiated and eventually interferes with the fields scattered there locally. This process to some extent resembles the operation principle of the plasmonic interferometers [55, 56], although in the case at hand it also physically expands the area that contributes to resonant scattering. Strong local response plays a crucial role here – it selects the actual mode that mediates non-local scattering. As we explained earlier, at the resonance of the zigzag metasurfaces the period of the plasmon waves has to fit inside one unit cell exactly.[§] This ensures that only λ/2-mode is sustained by the zigzag channels and all the unit cells engaged in non-local scattering will be driven to the corresponding λ/2-resonance and, therefore, radiate most strongly. More importantly, this guarantees that the radiated fields are all in phase and, hence, add up to form planar wavefronts, which stretch wide over the metasurface, significantly increasing the spatial coherence of the scattered light (see Fig. 4c). Nothing of this kind happens outside the resonance, as is evident from Fig. 4d.

The transmitted field then arises as a superposition of the planar wavefronts propagating in the forward direction, which are *sourced* by different unit cells exposed to light. In the case of incoherent illumination the exposed unit cells are excited with random phases and, hence, the planar wavefronts they produce do not interfere, preventing the transmission of a ZZnS metasurface from reaching its maximum. This is evident from the expression for transmitted light intensity written down below in terms of the planar waves scattered non-locally:

$$I^{tr} = \frac{c\varepsilon_0}{2} \left( \sum_{i=1}^{N} \left|E_i^{sca}\right|^2 + 2\,\text{Re}\sum_{i=1}^{N} \sum_{\substack{j=1 \\ (j \neq i)}}^{N} E_i^{sca} E_j^{sca*} \right), \qquad (1)$$

where $N$ is the number of unit cells, which are engaged in non-local scattering by each locally excited unit cell, and $E_i^{sca}$ is the complex amplitude of the planar wave *sourced* by $i$-th unit cell through non-local scattering. Given that the phases of $E_i^{sca}$ are random, all cross-terms in Eq. (1) cancel out. Correspondingly, under coherent illumination all unit cells oscillate in sync with exactly the same phase, which ensures that the cross-terms in Eq. (1) all add up (constructive interference of the scattered planar wavefronts) and, as a result, maximal transmission is achieved.

Besides providing a qualitative explanation of how the coherence of incident light controls the transmission of a ZZnS metasurface, Eq. (1) can be used to quickly estimate the strength of the effect. Assuming that the magnitudes of the scattered plane waves are equal (i.e., $\left|E_i^{sca}\right| = E$), the ratio between the levels of transmission for incoherent and coherent illumination $T_{inc}/T_c$ is simply given by $I_{inc}^{tr}/I_c^{tr} = E^2 N / E^2 N^2 = 1/N$. If incoherent illumination becomes partially coherent the ratio will need to be adjusted as follows: $T_{inc}/T_c = N_c/N$, where $N_c$ ($N_c \leq N$) is the number of unit cells that can fit within the coherence length, $d_c$, and hence will be excited with strong phase correlation. For practical use, it is convenient to express this ratio in terms of more accessible parameters:

$$\frac{T_{inc}}{T_c} = \frac{d_c}{L_{nl}\sqrt{2}}, \qquad (2)$$

where $L_{nl}$ characterizes the extent of non-local scattering and is equal to the maximum distance that an excitation can travel along the zigzags. The upper boundary for $L_{nl}$ is defined by the plasmon propagation length, which for gold nanostrips does not exceed 10 μm [57]. Correspondingly, for a ZZnS metasurface at $\lambda$ = 1.1 μm ($d_c$ = 2.4 μm), $T_{inc}/T_c \approx 0.2$. Given the above rather simplified assumptions, which did not take into account the variation of $E_i^{sca}$ along the metasurface and the shortening of the plasmon propagation length due to radiation loss, our estimate for $T_{inc}/T_c$ is not too far off the actual ratio of 0.3 measured in the experiment.

---

[§] In the near-IR the dispersion of surface plasmons in gold is still close to linear and therefore their wavelength can be roughly estimated as $\lambda_g \approx \lambda / \sqrt{(1+\varepsilon_s)/2}$, where $\varepsilon_s$ is the dielectric constant of the substrate [32]. Correspondingly, at the resonance $\lambda_g/2 = 0.44$ μm (which, given the assumption, is indeed very close the physical length of straight segments, $0.45$ μm).

In the case of a ZZnW metasurface the scattering process at the resonance is very similar to that revealed by Fig. 4c, producing (for every light beamlet) planar wavefronts stretching wide along the zigzags. The only differences are that here non-local scattering is mediated by the plasmonic mode of the nanowires, while the field scattered forward propagates alongside the field of the incident wave (which was effectively reflected back in the case of a ZZnS metasurface). Correspondingly, one may apply the above analysis to the field scattered backward by the ZZnW metasurface and show that non-local scattering controls reflection of the metasurface in exactly the same way as it controls transmission of the complimentary ZZnS design (in agreement with Babinet's principle), namely it reduces the intensity of light reflected by the ZZnW metasurface under incoherent illumination. This naturally translates into an enhancement of transmission occurring at the nanostructure's resonance, just as it was observed in our experiment with an incandescent while-light source (Fig. 3a).

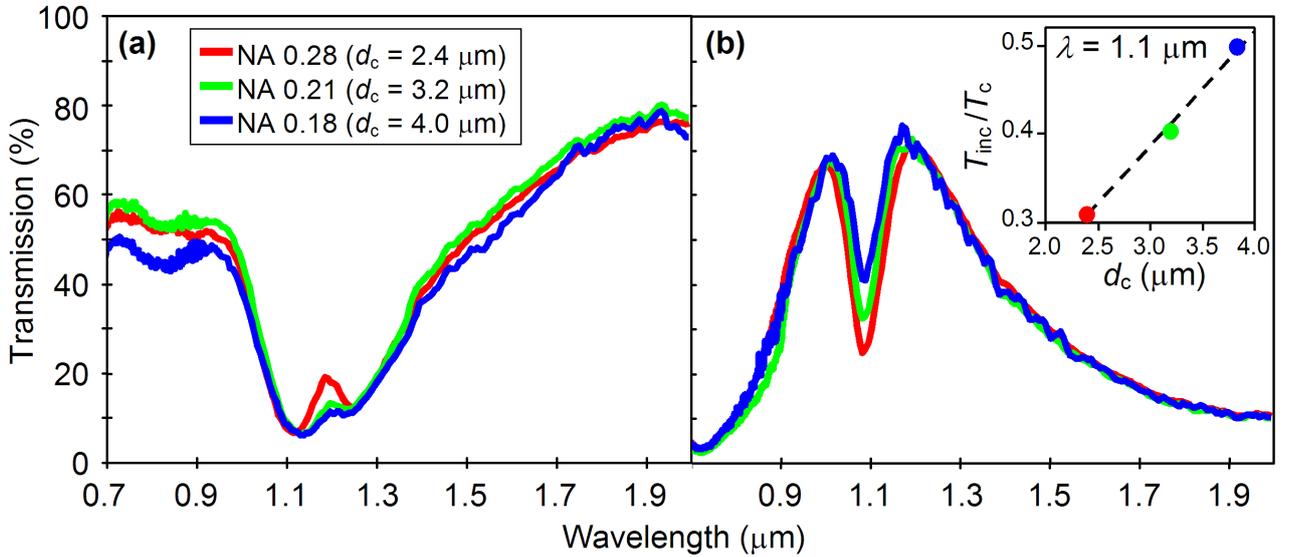

**Figure 5. Transmission response of zigzag metasurfaces under partially coherent illumination.** Panels (a) and (b) display transmission spectra of correspondingly ZZnW and ZZnS metasurfaces, which were acquired experimentally for three different values of the coherence length $d_c$. Inset to panel (b) plots $T_{inc}/T_c$ as a function of $d_c$ calculated for ZZnS metasurface using its transmission data in the main panel (coloured circles), and predicted by Eq. (2) based on the value of $T_{inc}/T_c$ at $d_c = 2.4$ μm (black dashed line).

Although the discovered effect may appear similar to the resonances of photonic crystal (PhC) slabs and the extraordinary optical transmission (EOT) of perforated metal films (which are well-known manifestations of strongly non-local response), there are several important differences to note. The new effect does not involve diffraction and is not possible to observe in the absence of strong (i.e. resonant) local response. Consequently, while PhC resonances and EOT should thrive on the spatial coherence of incident light (given their purely non-local nature), the new effect emerges only under spatially incoherent illumination and will gradually diminish with increasing degree of light coherence. To confirm this behaviour experimentally we have measured the transmission spectra of the zigzag metasurfaces using the same near-IR microspectrophotometer as before but with a reduced effective numerical aperture. For this study we chose NA 0.21 and NA 0.18, which corresponded to the spatial coherence length of 3.2 μm and 3.8 μm, and were obtained by decreasing the diameter of the condenser's diaphragm by 2 and 4 times, respectively. The measured spectra are plotted in Fig. 5, where they are compared with the transmission data obtained earlier using NA 0.28 ($d_c = 2.4$ μm). Clearly, the split in the resonances of ZZnW (Fig. 5a) and ZZnS (Fig. 5b) metasurfaces is becoming less pronounced as the coherence length of light increases. Moreover, the inset to Fig. 5b confirms that for the ZZnS metasurface the ratio $T_{\text{inc}}/T_{\text{c}}$ scales linearly with $d_c$, as predicted by Eq. (2).

In summary, we show experimentally and confirm via rigorous numerical modelling that the optical response of metallic metasurfaces may vary with the spatial coherence of incident light. This peculiar behaviour, previously unseen in artificially engineered materials, is characteristic, in particular, to nanostructured metasurfaces based on a continuous zigzag pattern. The two variants of such metasurfaces imposed by the pattern, namely arrays of zigzag nanowires and zigzag nanoslits, exhibit different levels of transmission at their resonance, depending on the degree of coherence of incident light. More specifically, upon switching from laser to partially coherent incandescent illumination, the intensity of transmitted light increased 3-fold in the case of zigzag nanowires and decreased 3-fold in the case of zigzag nanoslits. The mechanism underpinning the effect involves far-field interference of light waves scattered non-locally by the zigzag metasurfaces, while at resonance. The strength and robust nature of the effect make such metasurfaces immediately suitable for optical metrology applications. In particular, combined with a photodetector, a zigzag metasurface represents a very simple and compact optical device that will enable a quick assessment of light coherence, though at a predefined wavelength, complementing well the recently proposed comprehensive approach based on plasmonic interferometers [55]. Other possible applications will rely on the ability of the zigzag metasurfaces to selectively transmit or block spatially incoherent light, and may include the enhancement of optical imaging, vision, detection and communications.